\documentstyle[aps,twocolumn,prl,epsf,floats]{revtex}
\begin{document}
\draft

\twocolumn[\hsize\textwidth\columnwidth\hsize\csname @twocolumnfalse\endcsname
\title{Direct Observation of Antiferro-Quadrupolar Ordering\\
-- Resonant X-ray Scattering Study of DyB$_{2}$C$_{2}$ --}
\author{K.Hirota$^{1,4}$, N.Oumi$^{1}$, T.Matsumura$^{1,4}$, H.Nakao$^{2,4}$,
Y.Wakabayashi$^{2,3}$, Y.Murakami$^{2,4}$,  and
Y.Endoh$^{1,4}$}
\address{$^1$Department of Physics, Tohoku University, Sendai, 980-8578 Japan}
\address{$^2$Photon Factory, Institute of Materials Structure Science, KEK,
Tsukuba 305-0801, Japan}
\address{$^3$Department of Physics, Keio University, Yokohama 223-8522, Japan}
\address{$^4$CREST, Japan Science and Technology Corp., Tsukuba 305-0047, Japan}
\date{\today}
\maketitle
\begin{abstract}
Antiferroquadrupolar (AFQ) ordering has been conjectured in several
rare-earth compounds to explain their anomalous magnetic properties.  No direct
evidences for AFQ ordering, however, have been reported.  Using the resonant
x-ray scattering technique near the Dy $L_{III}$ absorption edge, we have
succeeded in observing the AFQ order parameter in DyB$_{2}$C$_{2}$ and analyzing
the energy and polarization dependence.  Much weaker coupling between orbital
degrees of freedom and lattice in $4f$ electron systems than in $3d$ compounds
provides an ideal platform to study orbital interactions originated from
electronic mechanisms.   
\end{abstract}
\pacs{PACS numbers: 61.10.-i, 71.20.g, 75.25.+z}
]
\narrowtext

Magnetic ions in a highly symmetrical crystalline may have an orbital degeneracy
in the crystalline electric field (CEF) ground state.  With decreasing
temperature, this degeneracy becomes lifted by some interactions.  A typical
example is the cooperative Jahn-Teller (JT) distortion,\cite{Gehring_75} where
the {\em orbital degrees of freedom}, coupled with lattice distortion, gives
rise to a structural phase transition to lower crystalline symmetry.  Long
range orbital ordering (OO) thus driven was confirmed in KCu$_{2}$F$_{4}$ by
polarized neutron scattering\cite{Akimitsu_76} and in LaMnO$_{3}$ by resonant
x-ray  scattering.\cite{Murakami_98b}  Although OO is not necessarily
associated with the cooperative JT distortion as reported in
La$_{0.88}$Sr$_{0.12}$MnO$_{3}$,\cite{Endoh_99} coupling with other degrees of
freedom such as charge or lattice would remain in
$3d$ compounds.\cite{Murakami_98a,Paolasini_99}

In $4f$ electron systems with degenerated ground state, however, orbital degrees
of freedom may remain and undergo a phase transition without structural
distortion because of much weaker coupling between lattice and well-localized
$4f$ orbitals.  Such a possibility was first discussed in
cubic CeB$_{6}$ as long range ordering of electric quadrupole moments of $4f$
orbitals.\cite{Morin_90}   Quadrupole ordering is defined as a phenomenon that
$f$ electron charge distribution {\it which diagonalizes certain quadrupole
moments} orders spontaneously and spatially as lowering temperature.  In
ferroquadrupole (FQ) arrangement, aligned quadrupole moments uniformly distort
the lattice through a linear coupling between quadrupole moment $O_{\Gamma}$ at
the wave vector $q=0$ and strain $\epsilon_{\Gamma}$ in the same symmetry. 
Therefore, the order parameter can be obtained by measuring JT-like lattice
distortion.\cite{Luthi_80}  In antiferroquadrupole (AFQ) arrangement, however,
the AFQ order parameter at $q \neq 0$ does not linearly couple with the uniform
strain.  Thus, atomic displacement is not always expected, which makes it
extremely difficult to observe the order parameter. 

\begin{figure}
\centerline{\epsfxsize=3in\epsfbox{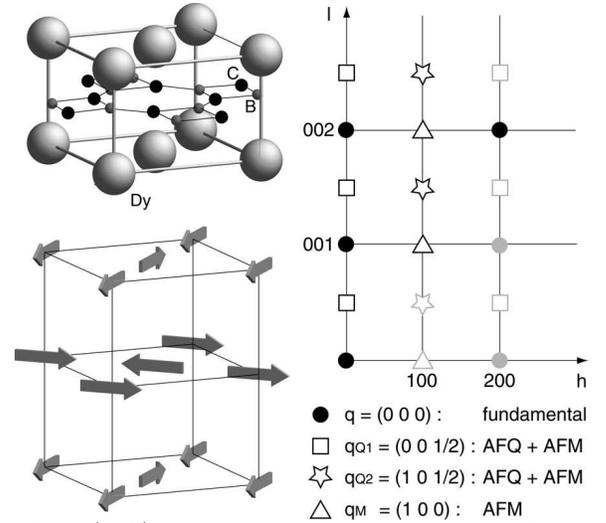}}
\caption{(Left) Crystal and magnetic structures of DyB$_{2}$C$_{2}$ ($P4/mbm$:
$a=5.341$~\AA, $c=3.547$~\AA\ at 30~K). (Right) Schematic view of $q$-space
investigated.}
\label{fig1}
\end{figure}

In this Letter, we present the first direct evidence for AFQ ordering by
the resonant x-ray scattering study of DyB$_{2}$C$_{2}$ near the Dy $L_{III}$
absorption edge.  As shown in Fig.~\ref{fig1}, DyB$_{2}$C$_{2}$ has the
$P4/mbm$ tetragonal structure consisting of Dy layers and B-C networks
stacking alternatively along the $c$ direction.\cite{Yamauchi_99}  Covalently
bonded B-C network requires no electron transfer from Dy$^{3+}$ ions ($4f^{9}$,
$^{6}H_{15/2}$), thus DyB$_{2}$C$_{2}$ has three conduction electrons per
formula in the 5$d$ band and is metallic.\cite{Bauer_80,Sakai_81}

%
%

Recently, Yamauchi {\it et al.}\cite{Yamauchi_99} reported that DyB$_{2}$C$_{2}$
exhibits phase transitions at $T_{Q} \sim 25$~K and $T_{N} \sim 16$~K.
Specific heat measurement showed two distinct $\lambda$-type anomalies at $T_{Q}$
and $T_{N}$, each of which releases the entropy equivalent to $R\ln 2$.  Since
Dy$^{3+}$ is a Kramers ion, two Kramers doublets should be involved
in these successive transitions.  It is thus expected that the ground and first
excited Kramers states are close {\it or} nearly degenerated and that quadrupole
degrees of freedom remain.  In contrast to the specific heat, almost no
anomaly was observed in the magnetic susceptibility at $T_{Q}$ and no structural
transition nor lattice distortion were confirmed at $T_{Q}$ and
$T_{N}$.  Thus, the transition at $T_{Q}$ is neither magnetic nor structural. 
Neutron diffraction revealed antiferromagnetic (AFM) ordering below
$T_{N}$.  Spins are aligned within the $c$ plane and the magnetic structure is
basically described with two propagation vectors $[1\ 0\ 0]$ and $[0\ 1\
\frac{1}{2}]$, indicating that Dy magnetic moments realize 90$^{\circ}$
arrangement along $c$, which is hardly explained only by magnetic
interactions.  They also found weak magnetic signals at $[0\ 0\ 0]$ and $[0\
0\ \frac{1}{2}]$, indicating that moments are slightly canted within the
$c$ plane. From these results, they proposed that the phase I ($T>T_{Q}$)
is paramagnetic, the phase II ($T_{N}<T<T_{Q}$) is the AFQ ordered phase, and
the phase III ($T<T_{N}$) is the AFM and AFQ ordered phase. 

We have grown a DyB$_{2}$C$_{2}$ single crystal by the Czochralski method.  The
crystal was checked by powder x-ray diffraction, which shows a diffraction
pattern consistent with Ref.~\onlinecite{Yamauchi_99} and no
detectable foreign phases.  The temperature dependence of magnetization is
also in good agreement.  X-ray scattering measurements were performed on a
six-axis diffractometer at the beamline 16A2 of the Photon Factory in KEK.  A
piece of the sample ($\sim 2$~mm cubic) was mounted in a closed cycle $^{4}$He
refrigerator so as to align the $c$-axis parallel to the $\phi$ axis of the
spectrometer.  The mosaicness was about 0.07$^{\circ}$ FWHM.  The azimuthal
angle $\Psi$ (rotation around the scattering vector) is defined as 0$^{\circ}$
where the scattering plane contains the $b$ axis, i.e., $[0\ 1\ 0]$.  The
incident energy was tuned near the Dy $L_{III}$ edge, which was experimentally
determined to be 7.792~keV using fluorescence.  To separate the linearly
polarized $\sigma'$ ($\perp$ the scattering plane)  and
$\pi'$ ($\parallel$ the scattering plane) components of diffracted beam, we used
the PG (006) reflection, which scattering angle is about 91$^{\circ}$ at this
energy resulting in almost complete polarization:
the $\sigma-\pi'/\sigma-\sigma'$ intensity ratio at $(0\ 0\ 2)$ was less than
0.5~\%.  In our configuration, $(0\ 0\ 2)$ intensity at Dy $L_{III}$
for $\sigma-\sigma'$ is $\sim 2.5 \times 10^{6}$ counts per second (cps) when
the  ring current is 300~mA.

AFQ ordering will be directly observed by exploiting the sensitivity of x-ray
scattering to an anisotropic $f$ electron distribution.  In the present study,
we have utilized the ATS (anisotropic tensor of x-ray susceptibility)
technique, which was originally developed for detecting ``forbidden
reflections'' which appear due to the asphericity of atomic electron
density.\cite{Dmitrienko_83,Templetons}  The ATS reflections, which are usually
very small, would increase in the resonant x-ray scattering near an
absorption edge because the anomalous scattering factor, sensitive to an
anisotropic charge distribution, is dramatically enhanced.  This technique was
successfully applied to the OO phenomena in 3$d$
oxides.\cite{Murakami_98b,Endoh_99,Murakami_98a,Paolasini_99}  We thus tuned the
incident energy of x-rays at Dy $L_{III}$, where $2p_{3/2}
\rightarrow 5d_{5/2}$ dipole and $2p_{3/2} \rightarrow
4f_{7/2}$ quadrupole transitions are expected.    

Figure~\ref{fig1} shows a schematic view of investigated $q$-space.  Fundamental
reflections appear where $h+k=even$.  To look for the AFQ and AFM
order parameters, we made scans along $(0\ 0\ l)$, $(\frac{1}{2}\ 0\ l)$, $(1\ 0\
l)$, $(h\ 0\ 2)$, $(\frac{1}{2}\ \frac{1}{2}\ l)$, $(1\ 1\ l)$ and $(2\ 1\ l)$
at Dy $L_{III}$ and making no polarization analysis.  Scans at 30~K give only
fundamental reflections.  At 20~K ($< T_{Q}$), two kinds of superlattice
reflections appear; one is characterized by a propagation vector $q_{Q1}=(0\ 0\
\frac{1}{2})$ and the other is by $q_{Q2}=(1\ 0\ \frac{1}{2})$.  At 10~K ($<
T_{N}$), additional reflections appear at forbidden reflection  points which
propagation vector is $q_{M}=(1\ 0\ 0)$.  These temperature dependences suggest
that the $q_{Q1}$ and $q_{Q2}$ reflections correspond to the expected AFQ
ordering and that the $q_{M}$ peak is the AFM order parameter.

%
%
\begin{figure}
\centerline{\epsfxsize=3in\epsfbox{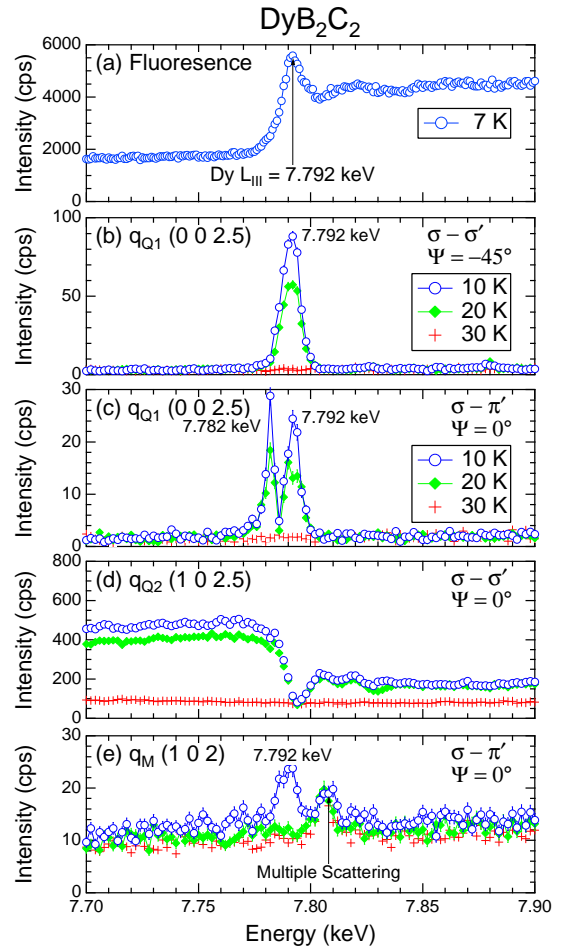}}
\caption{Incident energy dependences of (a) fluorescence (unpolarized), (b)
$q_{Q1}$ peak at $(0\ 0\ 2.5)$ for $\sigma-\sigma'$ polarization, (c) $(0\ 0\
2.5)$ for $\sigma-\pi'$, (d) $q_{Q2}$ peak at $(1\ 0\ 2.5)$ for
$\sigma-\sigma'$, and (e) $q_{M}$ peak at $(1\ 0\ 2)$ for $\sigma-\pi'$. No
background subtraction was made.}
\label{fig2}
\end{figure}

Figure~\ref{fig2} shows the incident energy dependences of fluorescence as
well as $(0\ 0\ 2.5)$,  $(1\ 0\ 2.5)$ and $(1\ 0\ 2)$ reflections, i.e,
$q_{Q1}$, $q_{Q2}$ and $q_{M}$ points.  The $(0\ 0\ 2.5)$ peak shows a sharp
enhancement  at Dy $L_{III}$ in both $\sigma-\sigma'$ and
$\sigma-\pi'$ processes.  Note that there exists another enhancement for
$\sigma-\pi'$ at 7.782~keV, 10~eV lower than the Dy $L_{III}$ edge, which we
speculate corresponds to level splitting within $2p$ and $5d$ states or a
quadrupole transition.  The $(1\ 0\ 2)$ peak shows an enhancement in
$\sigma-\pi'$ at the  Dy $L_{III}$ edge at 10~K.  No such enhancement was found
in $\sigma-\sigma'$ at $(1\ 0\ 2)$, indicating that the $(1\ 0\ 2)$ reflection
is dominated by $\sigma-\pi'$ scattering, as expected for resonant magnetic
scattering.  As for $(1\ 0\ 2.5)$, there exists a clear energy
enhancement in $\sigma-\pi'$ at Dy $L_{III}$ below $T_{N}$ indicating magnetic
contribution, which is consistent with Ref.~\onlinecite{Yamauchi_99}  On the
contrary, the $\sigma-\sigma'$ scattering exhibits {\em non-resonant}
reflection below $T_{Q}$ as shown in Fig.~\ref{fig2}(d).  We first focus upon
the resonant peaks, then discuss this non-resonant contribution at $(1\ 0\
2.5)$.

%
%
\begin{figure}
\centerline{\epsfxsize=3in\epsfbox{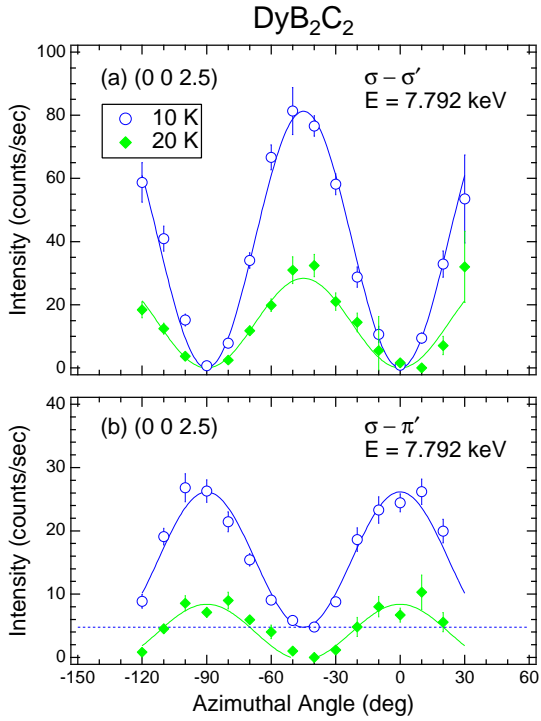}}
\caption{Azimuthal dependences of a $q_{Q1}$ peak $(0\ 0\ 2.5)$ for
$\sigma-\sigma'$ and $\sigma-\pi'$ polarizations.
$\omega-2\theta$ scan was made at each azimuthal angle after
lining up the crystal using $(0\ 0\ 2)$.  Intensity was obtained by fitting a
Gaussian to the peak profile.  The peak width of $(0\ 0\ 2.5)$ shows no
particular $\Psi$ dependence and no difference from that of
$(0\ 0\ 2)$, i.e, the instrumental resolution.  The solid curve is proportional 
to $\sin^{2} 2\Psi$.}
\label{fig3}
\end{figure}

In addition to the enhancement, it is expected that the resonant ATS scattering
from AFQ ordering shows the azimuthal angle dependence reflecting the
shape of $f$ electron distribution. As shown in Fig.~\ref{fig3}, we measured
azimuthal dependence for two different polarizations by rotating the crystal
around the scattering vector kept at $(0\ 0\ 2.5)$.  Figure~\ref{fig3}
demonstrates that the $\sigma-\sigma'$ scattering exhibits a characteristic
four-fold oscillation, compatible to the tetragonal symmetry.  The intensity
approaches zero at $\Psi=0$ and $\frac{\pi}{2}$.  The $\sigma-\pi'$ scattering
of  $(0\ 0\ 2.5)$ also shows  a four-fold oscillation.  However, the oscillation
for $\sigma-\pi'$ is reversed to that of $\sigma-\sigma'$. 
Plus, the intensity minimum remains finite at 10~K and approaches zero at
20~K, indicating that there exists a magnetic contribution to the
$\sigma-\pi'$ scattering at $(0\ 0\ 2.5)$, which is
consistent with Ref.~\onlinecite{Yamauchi_99}  These azimuthal dependences
strongly indicate the existence of anisotropic $f$ electron distribution and
the associated AFQ ordering below $T_{Q}$.

%
%
\begin{figure}
\centerline{\epsfxsize=3in\epsfbox{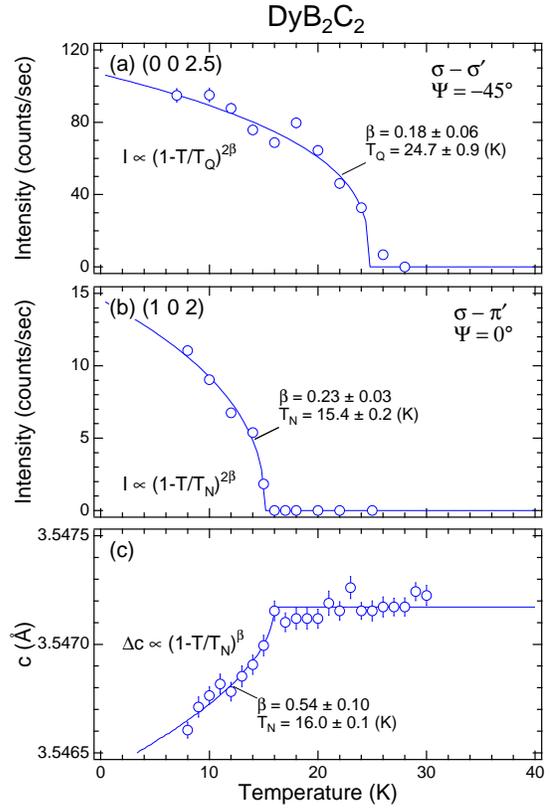}}
\caption{Temperature dependences of (a) AFQ and (b) AFM order parameters. Peak
profiles were obtained at particular polarization and azimuthal conditions so as
to maximize the intensities for $(0\ 0\ 2.5)$ and $(1\ 0\ 2)$, and fitted well
to a Gaussian.   Peak widths are resolution limited in the temperature range
where peaks are visible in our counting statistics (10 -- 15~seconds per
point).  (c) Temperature dependence of spontaneous strain $\Delta c$ estimated
using $(0\ 0\ 2)$.}
\label{fig4}
\end{figure}

Figures~\ref{fig4} show the order parameters measured at $(0\ 0\ 2.5)$ and $(1\
0\ 2)$ as well as the spontaneous strain $\Delta c$ estimated from the $(0\ 0\
2)$ peak position.  The order parameters behave as continuous 2nd
order transitions and can be fitted to power laws indicated in the figures. 
The transition temperatures thus obtained are in good agreement with the values
reported by Yamauchi {\it et al.}\cite{Yamauchi_99}  The critical exponents
$\beta$ obtained for the AFQ ordering and AFM ordering are about 0.2.  The
spontaneous strain $\Delta c$ has the $\beta$ value close to 0.5, which is twice
as much as that of the AFM ordering, indicating that $\Delta c$ is a secondary
order parameter and quadratically coupled to the AFM ordering.  For
quantitative discussions, we need more statistics, which will not only
give more precise $\beta$ values but also provide more information such as
correlation lengths above $T_{N}$ and $T_{Q}$.


Note that no anomaly was found in $\Delta c$ at $T_{Q}$, which implies that the
quadrupole ordering has very weak coupling, if any, to the lattice of
DyB$_{2}$C$_{2}$.  Since the superlattice peak at $(1\ 0\ 2.5)$ appearing
below $T_{Q}$ is non-resonant and has $\sigma-\sigma'$ polarization, it might
be ascribed to atomic displacement.  As a simple model for order estimation, let
us assume that Dy ions are displaced along $c$ and that the directions are
alternated between nearest neighbors.  From the intensity ratio below Dy
$L_{III}$, $I(1\ 0\ 2.5)/I(0\ 0\ 2) = 4.0 \times 10^{-5}$, we obtain the
displacement $\delta = 0.0014$~\AA\ (0.00040~$c$).  With this small atomic
displacement, the change of lattice constant may not be detected by the present
x-ray diffraction which resolution is $\Delta c/c \sim 10^{-4}$ (see
Fig.~\ref{fig4}(c)).  Recently, Benfatto {\it et al.}\cite{Benfatto_99}
theoretically reexamined the resonant x-ray scattering study of
LaMnO$_{3}$\cite{Murakami_98b} and argued that the resonant signal is mostly
due to the JT distortion resulting in anisotropic Mn-O bond lengths.  This is
in contrast with another theoretical description by Ishihara and
Maekawa,\cite{Ishihara_98} who proposed a mechanism based upon the Coulomb
interaction between $4p$ conduction band and the ordered $3d$ orbitals.  In
DyB$_{2}$C$_{2}$, however, it is very unlikely that the lattice distortion
results in the observed anisotropic electron distribution.  It is still 
required to further study the $(1\ 0\ 2.5)$ reflection, including the azimuthal
dependence, and to consider other possibilities such as asphericity of
atomic electron density due to AFQ ordering.


Let us briefly discuss the CEF of DyB$_{2}$C$_{2}$.  Using the equivalent
operator formalism,\cite{Stevens_52,Hutchings_53} we have constructed a point
charge model, which shows that the ground
($J_{z} = \pm \frac{1}{2}$) and first excited ($J_{z} = \pm \frac{3}{2}$)
Kramers doublets almost degenerate and are well separated from the other
excited states.  These results confirm the existence of a pseudo quartet ground
state in which the orbital degrees of freedom remain, and are consistent with
a strong planar magnetic anisotropy which aligns the magnetic moments within
the $c$ plane.\cite{Yamauchi_99}  Details of the calculation will be published
elsewhere.\cite{Matsumura_99}.

The present study unambiguously shows that the resonant scattering at $q_{Q1}$
corresponds to the AFQ ordering.  However, the mechanism yielding such resonant
scattering is not completely understood.  When allowed, the dipole transition
usually overwhelms the quadrupole transition in resonant scattering.  Similar
to a $d$ orbital angular moment, a quadrupole moment has five elements, i.e,
$Q_{m}^{(2)}$ ($m=2,1,0,-1,-2$) where $Q_{m}^{(l)}=\int
\rho({\bf r}){\bf r}^{l}\sqrt {4\pi/(2l+1)} Y_{lm} (\theta,\phi) d{\bf r}$ in
the polar coordinate.  In the CEF, the five elements are classified in
a particular irreducible representation, which can be conveniently explained by
the Stevens's equivalent operators\cite{Stevens_52}.  In the
cubic $O_{h}$ symmetry, for example, they are proportional to
$O_{2}^{0}=\{3J_{z}^{2}-J(J+1)\}/\sqrt{3}$ and
$O_{2}^{2}=J_{x}^{2}-J_{y}^{2}$ in the $\Gamma_{3}$ ($e_{g}$) symmetry, and
$O_{xy}=J_{x}J_{y}+J_{y}J_{x}$, $O_{yz}$, and $O_{zx}$ in the $\Gamma_{5}$
($t_{2g}$) symmetry.  Actual quadrupole moments can be obtained by
calculating their expected values.  Through a strong $c-f$
coupling between $5d$ conduction band and localized $4f$ orbitals, the
AFQ ordering would be projected onto the $5d$ orbital states.  To
completely understand the present experimental results, it is necessary to
establish much more detailed scattering mechanism in a proper CEF symmetry.

In conclusion, the present resonant ATS x-ray scattering study has directly
shown the existence of long range AFQ ordering in DyB$_{2}$C$_{2}$, which had
been theoretically conjectured in some $f$ electron systems, and given the
order parameter and the information concerning the final polarization and
azimuthal dependence, which are directly linked to the type of AFQ moments.

The authors are indebt to H. Yamauchi, H. Onodera and Y. Yamaguchi for
sharing their experimental results and sample preparation techniques prior to
publication.  We also acknowledge N. Kimura for helping single crystal growth
and T. Arima and S. Ishihara for crucial discussions.  This work was supported by
Core Research for Evolutional Science and Technology (CREST).

\vfill

\end{document}